\documentclass[aip,
 amsmath,amssymb,
 reprint,%
]{revtex4-1}

\usepackage{graphicx}% Include figure files
\usepackage{dcolumn}% Align table columns on decimal point
\usepackage{bm}% bold math
\usepackage[utf8]{inputenc}
\usepackage[T1]{fontenc}
\usepackage{mathptmx}
\usepackage{textcomp} % for \textsuperscript, \textmu and many more
\usepackage{nicefrac}
\usepackage{xcolor}
\usepackage{siunitx} % e.g. \si{\micro\joule}

\begin{document}

% \preprint{AIP/123-QED}

%\title{Time- and element-resolved magneto-optics using a high-repetition-rate fiber-laser-driven extreme ultraviolet high-harmonic generation light source}

\title{Ultrafast element-resolved magneto-optics using a fiber-laser-driven extreme ultraviolet light source}
% \title{Time- and element-resolved magneto-optics using a high-repetition-rate extreme ultraviolet high-harmonic light source}

\author{Christina Möller} % 1
\author{Henrike Probst} % 2
\author{Johannes Otto}\thanks{current address: IV. Physikalisches Institut, Universit\"at G\"ottingen, Friedrich-Hund-Platz 1, 37077 G\"ottingen, Germany} % 3
\author{Karen Stroh} % 4
\author{Carsten Mahn} % 5
\author{Sabine Steil} % 6
%\author{Manfred Albrecht} % ?
\author{Vasily Moshnyaga} % 7
\author{G.~S.~Matthijs Jansen} % 8
\author{Daniel Steil} % 9
\author{Stefan Mathias} \email{smathias@uni-goettingen.de}% 10
\address{I. Physikalisches Institut, Universit\"at G\"ottingen, Friedrich-Hund-Platz 1, 37077 G\"ottingen, Germany}
\address{International Center for Advanced Studies of Energy Conversion (ICASEC), Georg-August-Universit\"at G\"ottingen, 37077, G\"ottingen, Germany}

\begin{abstract}
We present a novel setup to measure the transverse magneto-optical Kerr effect in the extreme ultraviolet spectral range at exceptionally high repetition rates based on a fiber laser amplifier system. This affords a very high and stable flux of extreme ultraviolet light, which we use to measure element-resolved demagnetization dynamics with unprecedented depth of information. Furthermore, the setup is equipped with a strong electromagnet and a cryostat, allowing measurements between 10 and 420~K using magnetic fields up to 0.86 T. The performance of our setup is demonstrated by a set of temperature- and time-dependent magnetization measurements showing distinct element-dependent behavior.
\end{abstract}

\maketitle
\section{\label{sec:intro} Introduction}
In the last decades, magneto-optical spectroscopy has lead to an impressive series of discoveries in ultrafast magnetism, spanning from the discovery of optically-induced ultrafast manipulation of magnetic order\cite{beaurepaire1996ultrafast, RevModPhys2010} to the experimental verification of the theoretically-predicted existence of femtosecond spin currents\cite{Battiato2010, Rudolf2012, Turgut2013} and optically-induced spin transfer\cite{Elliott2016,Dewhurst2018,Siegrist2019, Hofherr2020, Willems2020, Chen2019, Tengdin2020, steil_efficiency_2020}. Nevertheless, a complete and thorough understanding of the phenomena on a microscopic scale is still an area of active research. Here, the use of ultrashort extreme ultraviolet (EUV) light pulses to probe the instantaneous magnetization enables both few femtosecond time-resolution and the possibility to disentangle the contributions of individual elemental components of the probed system\cite{radu_transient_2011, La-o-vorakiat2012}. Making use of the transverse magneto-optical Kerr effect (T-MOKE), full time- as well as energy-resolved (and thus element-resolved) data can be acquired from the wavelength-dependent reflection of the sample\cite{La-o-vokariat2009,Hofherr2020, jana_analysis_2020}. 

A commonly used light source for both magnetic\cite{La-o-vokariat2009, Jana2017, yao_tabletop_2020, Siegrist2019} and non-magnetic\cite{geneaux_transient_2019} extreme ultraviolet spectroscopy is high-harmonic generation (HHG)\cite{lewenstein_theory_1994, popmintchev_attosecond_2010}. HHG combines a compact, laboratory-scale light source with excellent properties of the EUV light; namely a high degree of coherence, broad bandwidth and ultrashort pulse durations in the low-femtosecond to attosecond regime, allowing state-of-the-art time resolution in pump-probe experiments. In comparison to alternative short-pulse EUV light sources such as free-electron lasers\cite{ackermann_operation_2007} and femtoslicing synchrotrons\cite{schoenlein_generation_2000, zholents_femtosecond_1996}, generation of a stable EUV source with high average power is challenging. In that regard, several approaches based on high-repetition-rate, high-power lasers have recently been developed\cite{boutu_overview_2015, hadrich_single-pass_2016}. Up until now however, laboratory-based extreme ultraviolet magneto-optic experiments have been limited by the use of Ti:Sapphire laser systems at repetition rates below a few ten kilohertz\cite{La-o-vokariat2009, Jana2017, jana_analysis_2020, yao_tabletop_2020, Siegrist2019}.

In this article, we present a novel EUV T-MOKE experiment providing massively improved signal quality due to the use of a high-repetition-rate fiber laser system to drive the high-harmonic generation. Amongst others, our photon energy range of 30 to 72~eV covers the M\textsubscript{2,3} absorption edges of the \textit{3d} transition metals Co, Ni, Fe and Mn, allowing a simultaneous probing of magnetic properties of these elements. In contrast to earlier HHG-based femtomagnetism experiments, we operate at much higher repetition rates up to 300~kHz, which allows for extremely high signal quality in only limited acquisition time. Furthermore, we are able to apply high magnetic fields of up to 0.86~T and temperatures between 10~K and 420~K in our setup. 

\section{\label{sec:EUV}Setup for time-resolved T-MOKE at EUV wavelengths}

\begin{figure*}[htb]
\centering
\includegraphics[width=\textwidth]{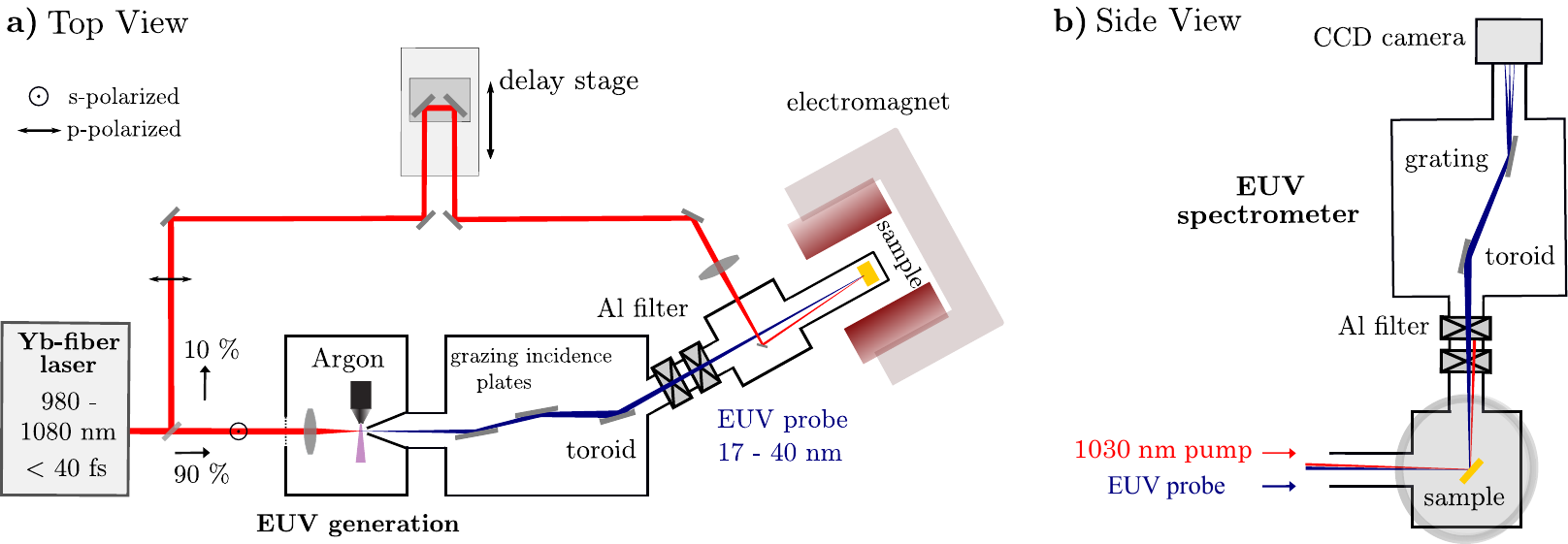}
\caption{Schematic overview of the setup: a) top view: High harmonics are generated in an argon gas jet by intense IR pulses at a repetition rate up to 300 kHz. The fundamental beam is consequently filtered out by two grazing incidence plates and an Al filter. The toroidal mirror focuses the EUV beam onto the sample inside the electromagnet.  b) side view: After reflection on the sample, the high harmonic spectrum is analyzed by our custom-built spectrometer composed of a toroidal mirror, a grating and a CCD camera. The pump beam is coupled into the vacuum chamber and blocked by a second Al filter after reflection on the sample.}
\label{fig:setup}
\end{figure*}

The experiment makes use of the transverse magneto-optical Kerr effect (T-MOKE) to measure the magnetization of the sample. In T-MOKE, the sample magnetization is aligned parallel to the sample surface and perpendicular to the plane of incidence of the light. Here, a magnetization-dependent change in the reflectivity can be observed for p-polarized light. As this can be measured by a simple intensity measurement, T-MOKE is especially well-suited in the EUV spectral range, where polarization analysis is non-trivial\cite{Yamamoto2017}.

The quantity accessible by T-MOKE is the magnetic asymmetry $A$, which is the difference in reflected intensity from the sample for both magnetization directions ($I_\uparrow$ and $I_\downarrow$), 
normalized to the sum of the intensities
\begin{align}
    A &= \frac{I_\uparrow - I_\downarrow}{I_\uparrow + I_\downarrow}
    \approx 2\text{Re}\left[\frac{\sin(2\theta_0)\epsilon_{xy}}{n^4\cos^2\theta_0-n^2+\sin^2\theta_0}\right].
    % 2\text{Re}\left[\frac{\sin(2\theta_0)\epsilon_{xy}}{n^4\cos^2\theta_0-\sin^2\theta_0(n^2+1)}\right]
    \label{asymmetry_equation}
\end{align}
The index of refraction $n$ and the off-diagonal elements of the dielectric tensor $\epsilon_{xy}$ describe the reaction of the full quantum-mechanical system to the incident light field, and can be calculated by considering all possible electronic transitions for the material's spin-resolved band structure and instantaneous electron occupation.\cite{yaresko_band-structure_2006, jana_analysis_2020} As a result, both $n$ and $\epsilon_{xy}$ are wavelength-dependent, and change rapidly near absorption edges. In particular, resonant enhancement of $\epsilon_{xy}$ at the M edges of transition metals\cite{hecker_soft_2005, pretorius_transverse_1997} combined with a general enhancement of the magnetic asymmetry at absorption edges\cite{La-o-vorakiat2012} leads to a strong, wavelength-dependent T-MOKE signal in the EUV range for the \textit{3d} transition metals. Here, spectroscopy can be optimally used to disentangle elemental contributions from the magnetic asymmetry. 

In the second step of Eq.~\ref{asymmetry_equation}, the magnetic asymmetry was approximated by inserting the expressions for $p$-polarized reflected intensities under the assumption that the Fresnel coefficient is large compared to the magneto-optical term.\cite{La-o-vorakiat2012} Higher order terms of 
the magnetization-dependent off-diagonal element of the dielectric tensor $\epsilon_{xy}$ are neglected. From this expression, it can be deduced that, in thermal equilibrium, the asymmetry is proportional to the magnetization inside the sample. The expression is maximized when the denominator approaches zero. This is exactly the case for $\tan\theta_0=n$, which is the condition for the Brewster angle. Since $n\approx 1$ in the EUV range, the optimal angle of incidence corresponds to $\theta_0=45$\textdegree{}. 

\subsection{The fiber-laser HHG light source}
Here, we present a T-MOKE setup using EUV light from a high-harmonic generation (HHG) light source to probe the sample. Ultrafast (de)magnetization dynamics are induced by tuneable pump pulses from the same high repetition-rate fiber laser system used to drive the HHG. After reflection from the sample, the EUV light is dispersed in a custom-built spectrometer and the diffracted harmonics are detected by a CCD camera. A schematic overview of the setup is shown in Fig.~\ref{fig:setup}.

Ultrashort laser pulses with a center wavelength of 1030~nm, tuneable repetition rate up to 300~kHz, pulse energy of up to 150~\si{\micro\joule} and pulse lengths below 40~fs are generated by a fiber laser system (Active Fiber Systems). This is achieved by self-phase modulation of 300~fs pulses from an ytterbium-doped fiber amplifier in a krypton-filled hollow-core fiber and subsequent compression by chirped mirrors. The beam is split into a pump and a probe beam by a variable attenuator, consisting of a $\lambda$/2-plate and two broadband thin film polarizers. 
The s-polarized output of the attenuator is focused by a 100~\si{\milli\meter} focal length lens into an argon gas jet to generate high harmonics, which provides the probe beam in our experiment. In order to optimize the HHG process, the gas jet is mounted on a 3-axis positioning system. Efficient HHG was observed at an argon backing pressure of 17~bar for a gas nozzle diameter of 100~\si{\micro\meter}.

To minimize reabsorption of the generated EUV beam, the chamber is pumped by a multi-stage root pump (Ebara EV-SA30) and a 1~\si{\milli\meter} differential pumping aperture is placed directly behind the generation point. This results in a pressure of $2\times10^{-4}$~mbar in the subsequent mirror chamber despite the quite high gas load in the first chamber (1.1~mbar).
The generated high harmonics are separated from the fundamental by two grazing-incidence plates (GIP) inside the mirror chamber.\cite{pronin_ultrabroadband_2011} These fused silica plates are anti-reflection coated for 1030~nm on the front and backside, whereby most of the fundamental beam is transmitted and dumped outside the vacuum chamber. An additional top Ta$_2$O$_5$ coating of the GIP allows for a high reflection of the EUV spectrum at 10\textdegree{} grazing incidence. After the GIPs, a B$_4$C-coated toroidal mirror (276~\si{\milli\meter} focal length at 15\textdegree{} grazing incidence angle) focuses the probe beam onto the sample in a 350x100~\si{\square\micro\meter} spot (1/$e^2$-diameter). The EUV light then passes a 100~nm thick aluminum filter which is mounted in front of the sample in a VAT valve to block any residual fundamental light. 

\subsection{Pump beamline}
A second variable attenuator in the pump beamline allows for manipulation of the pump beam intensity. A motorized delay line changes the optical path length with respect to the probe path. After passing a 350~\si{\milli\meter} focal length lens, the pump beam is coupled into the vacuum chamber such that the angle between the pump and the EUV beamline is less than $2^\circ$. 
The beam diameter at the sample position was determined to be 760x720~\si{\square\micro\meter}, yielding a pump fluence of 1.75~\si[per-mode=symbol]{\milli\joule\per\square\centi\meter} for an average power of 750~mW at a repetition rate of 100~kHz. The pulse duration is measured by intensity autocorrelation to be 58~fs, assuming a Gaussian pulse shape. The pump beam is s-polarized with respect to the plane of incidence on the sample for the measurements shown in this paper.

\begin{figure}[htb]
    \centering
    \includegraphics[width=\linewidth]{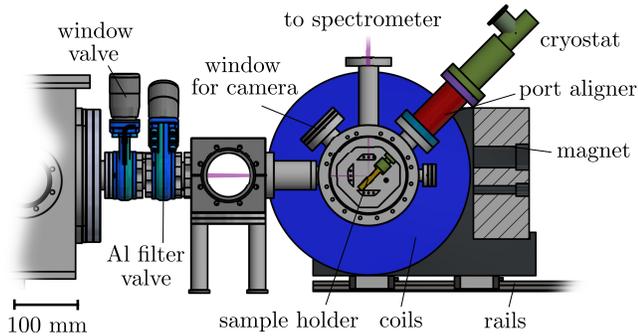}
\caption{Technical drawing of the custom-built sample chamber, providing a view directly into the sample chamber.}
\label{fig:sample-chamber}
\end{figure}

\subsection{Sample environment}
The sample environment is a custom-made chamber based on a CF150 flange and designed to fit in the 45~\si{\milli\meter} wide pole gap of the electromagnet (see Fig.~\ref{fig:sample-chamber}). The flange can be opened from the side, which enables easy access to the sample. For this purpose the electromagnet surrounding the sample chamber is mounted on rails, allowing to remove the magnet and providing unobstructed access to the sample chamber. The switching time of the electromagnet varies from 60~\si{\milli\second} for 100~\si{\milli\tesla} up to 600~\si{\milli\second} for the maximum field strength of 860~mT.

The samples are mounted on a removable copper sample holder attached to a cryostat, which is fixed at a 45\textdegree{} angle with respect to the incoming pump and probe beams. In order to make small changes to the position and angle of the sample, the cryostat is attached to the chamber with an ultrahigh vacuum (UHV) port aligner. The continuous-flow UHV cryostat (Janis Research ST400) enables temperature control of the sample in the range of $\approx$10 to 420~K. Typically, the pressure inside the chamber is below $1\times10^{-7}$~mbar during operation, however, even UHV conditions can be achieved by baking out the vacuum chamber. 

To verify spatial overlap of the pump and probe beam, the diffuse reflection of the pump and fundamental of the high harmonics on the sample is imaged with a CMOS sensor. The temporal overlap is achieved by observing sum frequency generation of both IR pulses in a beta barium borate (BBO) crystal at the sample position or recording interference fringes with a beam profiler. Both of these methods can be performed outside of the vacuum system by inserting a moveable mirror directly in front of the sample. This determination of the temporal overlap is limited by a small difference in the optical path lengths of the fundamental and the high harmonics. Therefore, a final determination of the temporal overlap is done by searching for the onset of demagnetization as probed by the EUV light.

\subsection{Spectrometer design}
Spectroscopy of the reflected EUV light allows for the investigation of the sample magnetization and is realized by a custom-built spectrometer based on a plane grating as the dispersive element and a toroidal mirror as the focusing element. The spectrometer was designed to achieve a high energy resolution to distinguish between the individual harmonics and to provide in-detail access to spectrally distinct dynamics around the absorption edges of the different elements. 

The sample reflects both the pump and probe beams towards the spectrometer chamber (see Fig.~\ref{fig:setup}b). Here, a second aluminum filter is introduced into the beam path to block the reflected pump beam. The EUV beam is focused towards the CCD-camera (GE 2048 512 BI UV, Greateyes) by a toroidal mirror, which is identical to the toroidal mirror used to focus the EUV probe beam on the sample. 
The planar reflection grating is positioned in the off-plane geometry, where the grating grooves lie parallel to the plane of incidence of the light. 
An advantage of this geometry is that it allows for the use of efficient blazed gratings while maintaining the low grazing incidence angle which is necessary for high reflectivities at EUV wavelengths~\cite{frassetto_single-grating_2011}.
The grating is gold-coated and has a groove density of 1800~lines/mm. Given the blaze angle $\delta$ of 9.3\textdegree{}, optimal diffraction efficiency is achieved at the altitude angle $\gamma$ of 6.5\textdegree{}  for wavelengths near the M-edges of the \textit{3d} ferromagnets.
To prevent camera exposure during the sensor readout procedure, a shutter is positioned between the toroidal mirror and the grating.  

\begin{figure*}[htb!]
\centering
\includegraphics[width=\textwidth]{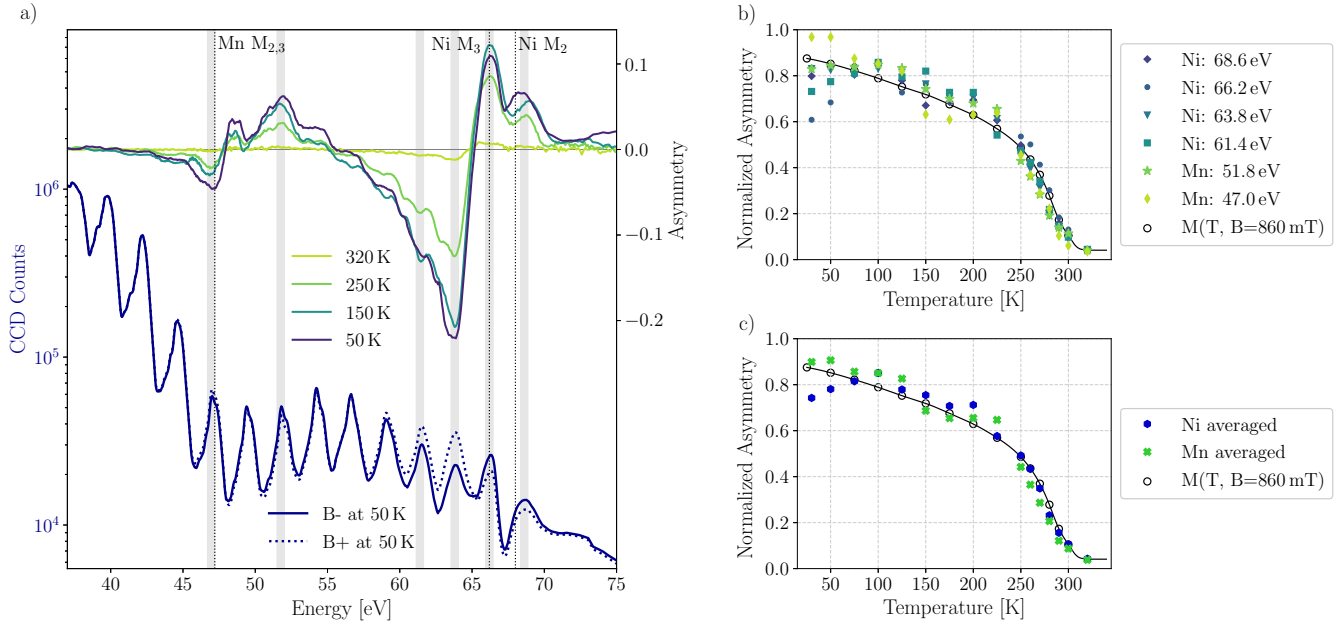}
\caption{Temperature dependence of the magnetic asymmetry of the LNMO phase transition. a) The magnetic asymmetry, together with typical HHG spectra at 50~K for the different magnetization directions. b) Temperature-dependent magnetic asymmetry for 0.5~eV energy intervals around the relevant high harmonic energies of Mn and Ni. c) The magnetic asymmetry traces for Mn and Ni, averaged over all relevant high harmonics. The asymmetry traces in (b) and (c) nicely follow the $M(T,B=860~\si{mT})$ data points which were recorded using SQUID magnetometry. The black line is an interpolation of the $M(T,B=860~\si{mT})$ data and serves as a guide to the eye. }
\label{fig:LNMO_M_T}
\end{figure*}

A consequence of this spectrometer design is that the EUV wavefront incident on the grating is curved. Equivalently, it can be said that the azimuth angle $\theta$ and the altitude angle $\gamma$ of the beam on the grating have a certain spread. This will lead to a distortion of the measured spectrum, which is calculated here. The grating equation for this system is\cite{cash_echelle_1982} 
\begin{equation}
    m \lambda \sigma = \sin \gamma (\sin \theta - \sin \theta'),
\end{equation}
where $m$ is the diffraction order, $\lambda$ the wavelength, $\sigma$ the grating constant and $\theta'$ the exit angle. Substituting $\theta \rightarrow \theta_0 + \Delta \theta$ and $\gamma \rightarrow \gamma_0 + \Delta \gamma$ and performing a Taylor expansion around $\Delta \theta = \Delta \gamma = 0$, we find
\begin{equation}
\label{eq:taylor}
    \theta' = \theta'_0 - c_1 \Delta \theta + c_2 \Delta \gamma + O(\Delta \theta ^2+ \Delta \theta \Delta \gamma + \Delta \gamma ^2).
\end{equation}
In our experimental geometry, $c_1 \approx 1.1$ and $c_2$ is typically on the order of 3. Higher order terms can be ignored because they are typically one to two orders of magnitude weaker than the linear terms. From this result, we can conclude the following: 1) the term linear in $\Delta \theta$ will increase the wavefront curvature, leading to an astigmatic focus of the diffracted beam. On the camera, this results in spots which are elongated perpendicular to the spectral direction. 2) the term linear in $\Delta \gamma$ will shift the diffracted beams depending on their altitude angle. Effectively, this leads to a grating dispersion which is stronger for shallower grazing incidence angles on the grating.

In order to correct for this deformation of the measured spectrum, we first transform the cartesian coordinates of the camera onto a ($\theta$, $\gamma$)-coordinate system. Next, we apply a $\gamma$-dependent scaling of the diffraction angle, eliminating the tilt of the high harmonics. Finally, we integrate the two-dimensional measurement along the spatial dimension without loss of spectral resolution. 

In daily operation, the aluminum L$_3$ edge at 72.7 eV and the known separation of the high harmonics by twice the fundamental frequency are used to calibrate the energy of the recorded high harmonics. In addition, the energy calibration of the spectrum is verified by inserting a 200~nm thick tin filter, which transmits EUV light up to 23.6~eV allowing to identify the 19\textsuperscript{th} harmonic (22.9~eV) of our fundamental driver.

\section{\label{sec:static}Exemplary experimental results and analysis}

In the following, we present temperature- and time-dependent measurements of the magnetic asymmetry. The results show the ferromagnetic phase transition of La$_2$NiMnO$_6$ and the distinct ultrafast magnetization dynamics of iron and nickel in a Fe$_{19}$Ni$_{81}$ alloy. These measurements demonstrate the capabilities of our setup to record magnetization traces with a high signal-to-noise ratio over a broad range of sample temperatures between 10 and 420 K and with femtosecond time resolution.

\subsection{Temperature-dependent magnetic asymmetry}

% \begin{figure*}[hbt!]
% \centering
% \includegraphics[width=\textwidth]{Fe19Ni81_1030nmPump_M28_20201124_750mW_50_7to54_5eV_63_4to67_5eV_new_fit_58fs_gauss.eps}
% \caption{Time-resolved demagnetization measurement of Fe\textsubscript{19}Ni\textsubscript{81}. 
% a) Typical HHG spectra for the different magnetization directions before the onset of demagnetization and resulting magnetic asymmetry. The dashed lines depict the known absorption edge positions. 
% b) Relative demagnetization time traces for iron and nickel. Here, the asymmetry is averaged over the depicted energy intervals in (a) and normalized to time delays before time zero. The shaded area indicates the standard deviation of individual data points.
% c) Zoom-in on the first 800~fs. The solid lines are fits according to a double exponential function (eq. \eqref{time_dependent}).
% }
% \label{fig:FeNi_tr}
% \end{figure*}

\begin{figure*}[hbt!]
\centering
\includegraphics[width=0.85\textwidth]{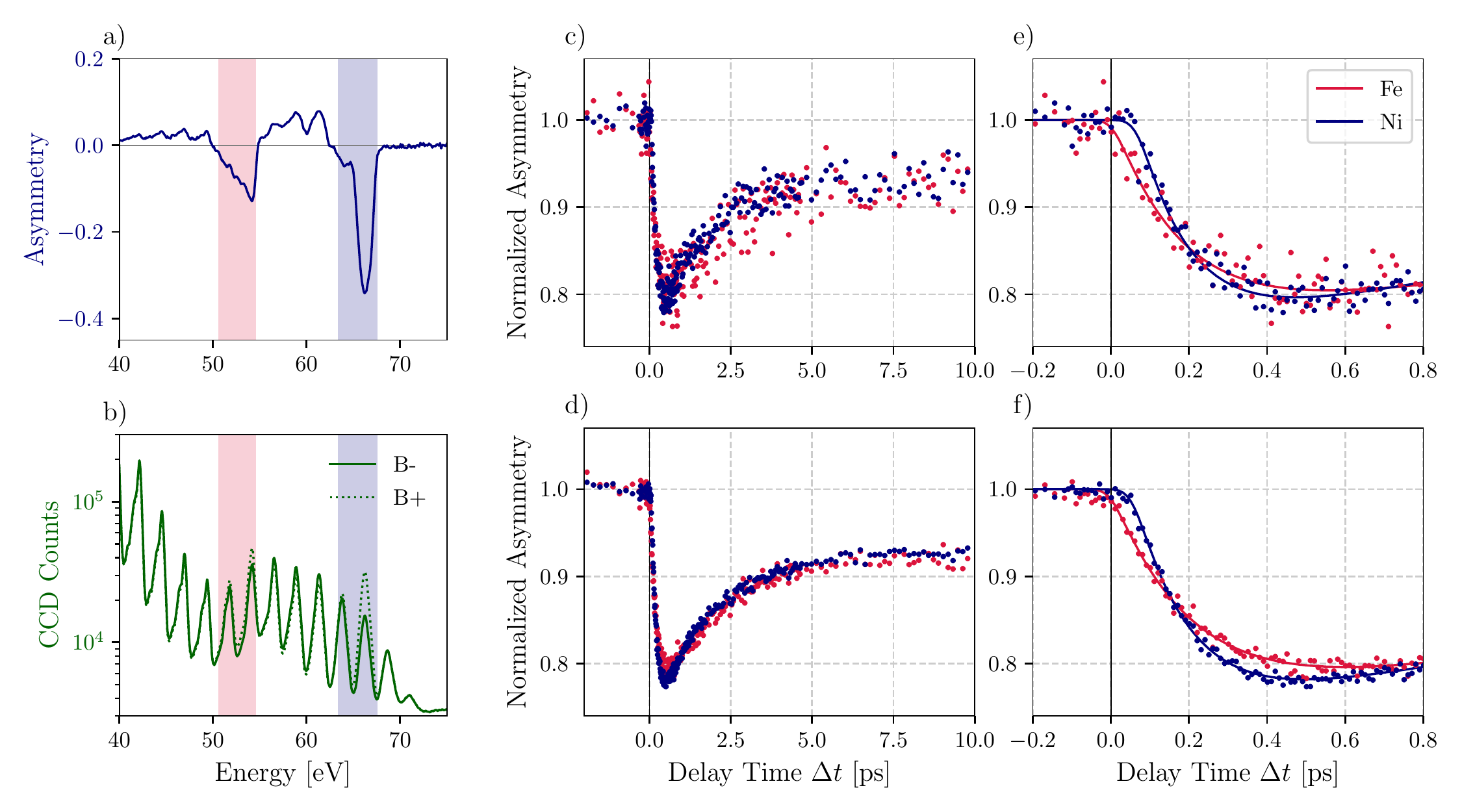}
\caption{Time-resolved demagnetization measurement of Fe\textsubscript{19}Ni\textsubscript{81} for a short (1 hour) and long (12 hours) time measurement. 
a) Typical magnetic asymmetry of the sample before arrival of the pump pulse.
b) Corresponding high-harmonic spectra for both magnetic field directions.
c, d) Relative demagnetization time traces for iron and nickel for one and twelve hours measurement time respectively. Here, the asymmetry is averaged over the depicted energy intervals in (a) and normalized to time delays before time zero. 
e, f) Zoom-in on the short-timescale dynamics of (c) and (d) respectively. The solid lines are least-square fits according to a double exponential function (eq.~\eqref{time_dependent}), the results of which are given in table~\ref{tab:fit_results_double}.
}
\label{fig:FeNi_tr}
\end{figure*}

Thin-film perovskite magnanites exhibit rich phase diagrams due to the strong correlations between charge, lattice and spin degrees of freedom~\cite{imada1998metal, tokura1999colossal, moshnyaga2007}, which can be of particular interest for ultrafast spin dynamics. Here, we investigated the ferromagnetic to paramagnetic phase transition in the double perovskite La$_2$NiMnO$_6$ (LNMO). %The 100~nm thick sample was prepared on MgO(100) by metalorganic aerosol deposition technique at high oxygen partial pressure, $p\text{O}_2  \approx 0.2$~bar.\\
The magnetic asymmetry was recorded in a temperature range of 30~K to 320~K. Typical HHG spectra at 50~K for both magnetization directions and the resulting asymmetry spanning the M\textsubscript{2,3} absorption edges of manganese (47.2~eV) and nickel (66.2~eV and 68~eV) for different temperatures are depicted in Fig.~\ref{fig:LNMO_M_T}a. The measurements were performed in an applied magnetic field of 860~mT. For each temperature, 500 spectra with an exposure time of 1 second per magnetization direction were recorded.
In order to analyze the quantitative temperature dependence of the asymmetry, we calculated the mean asymmetry in a 0.5~eV energy interval around the relevant high harmonic peaks for Mn and Ni, indicated by the shaded areas in Fig.~\ref{fig:LNMO_M_T}a.
As a reference, we extracted the sample magnetization at $B=860~\si{mT}$ from full hysteresis curves $M(B)$ in the temperature range of 5-300 K as measured using a superconducting quantum interference device (SQUID) magnetometer.
In order to compare the observed magnetic asymmetry to the $M(T,B=860~\si{mT})$ data, we have normalized the magnetic asymmetry to the sum of the asymmetry values for all temperatures and then scaled it with the magnetization $M(T,B=860~\si{mT})$ integrated over the full temperature range.

The temperature-dependent magnetic asymmetry for all relevant harmonic peaks is shown together with the $M(T,B=860~\si{mT})$ data in Fig.~\ref{fig:LNMO_M_T}b. The normalized asymmetry values for Mn and Ni show good agreement with the $M(T,B=860~\si{mT})$ reference measurement performed with SQUID across the entire temperature range. The expected loss of magnetization with increasing temperature is clearly visible.  %The normalized asymmetry values for Mn and Ni show good agreement with the $M(T,B=860~\si{mT})$ data across the entire temperature range. The loss of magnetization close to the Curie temperature $T_C=284$~K, as determined from a low-field M(T) SQUID measurement at $B=100~\si{mT}$, is clearly visible. 
The remaining non-negligible magnetization beyond the phase transition in Fig.~\ref{fig:LNMO_M_T}b is induced by the strong external field of 860~\si{\milli\tesla} applied during the measurement and evident in both the T-MOKE measurement and SQUID $M(T,B=860~\si{mT})$ data. 

To extract a mean value of the magnetic asymmetry for both Mn and Ni, we averaged the asymmetry of the relevant harmonics for both elements. The resulting averaged magnetic asymmetry traces are shown in Fig.~\ref{fig:LNMO_M_T}c. From the overall agreement between the EUV magnetic asymmetry measurement and the SQUID measurement, we conclude that static EUV T-MOKE effectively probes the bulk magnetization in the steady state. 
% Comparing the different spectral regions allows us to extract the magnetization behaviour of the nickel and manganese sublattices in the LMNO sample independently. At high temperatures, we tentatively observe a small difference in the Curie temperature of nickel and manganese, with manganese being roughly 10~\textdegree{}C lower. Also at the lowest temperatures we observe a difference, with a relative increase in the contribution of manganese. In this case however, further measurements are necessary to verify that these differences are not caused by the formation of a thin layer of ice on the sample.
%Qualitatively, we observe a less smoother phase transition, with significantly lower spin polarization than expected at temperatures between 220 and 280~K.

\subsection{\label{sec:trmoke} Time-resolved EUV-T-MOKE}

To demonstrate the performance of our setup we recorded the time evolution of the ultrafast demagnetization in a Fe$_{19}$Ni$_{81}$ thin film at 100~\si{\milli\tesla} applied field. Ultrafast demagnetization is initiated by a 58~fs, 1.2~eV pump pulse (measured by autocorrelation, assuming Gaussian pulse shape). The  absorbed pump fluence is $\approx$ 0.7~\si[per-mode=symbol]{\milli\joule \per\square\centi\meter}. At this fluence, the 700~mW thermal load on the sample leads to a limited thermal demagnetization of around $5~\%$ for our pulse repetition rate of 100~kHz. The measurement time was 12 hours which corresponds to 96 0.7-second measurements of the asymmetry per delay position.  We will compare the results from this long integration to results from shorter integration slices selected from this data.

The results are summarized in Fig.~\ref{fig:FeNi_tr}, including recorded EUV spectra for both magnetization directions after subtraction of a background image, the resulting asymmetry according to equation~\eqref{asymmetry_equation}, and the time dependence of the normalized magnetic asymmetry for a 12-hour measurement as well as an exemplary 1-hour data set selected from the long measurement. For the time-dependent data, the magnetic asymmetry was averaged over the indicated energy intervals in Fig.~\ref{fig:FeNi_tr}a. Negative time delays indicate that the EUV pulse probes the sample before the pump pulse arrives. %We note that in Fig.~\ref{fig:FeNi_tr}d, a small dynamic demagnetization can already be observed before the arrival of the main pump pulse. This can be attributed to a small pulse pedestal from the laser amplifier. This can easily be removed by first modifying the pump pulse in a non-linear optical process such as second harmonic generation, but it is not of great consequence for the results presented here.

For the analysis of the raw asymmetry data, large energy regions around the M absorption edges were chosen: Fe between 50.7 and 54.5~eV and Ni between 63.4 and 67.5~eV. By integrating the off-diagonal component of the dielectric tensor over a wider energy range, we are less sensitive to spectrally distinct band-structure renormalization and population dynamics\cite{Eich2017, Gort2020, Hofherr2020, Willems2020, jana_analysis_2020, Hennes2020, Yao2020}, and we effectively probe the element-specific magnetization. Already from the 1-hour dataset, we can easily resolve the delayed demagnetization dynamics between nickel and iron, as previously seen in Refs.~\onlinecite{Mathias2012, Jana2017, Guenther2014, Hofherr2020}, which highlights the excellent signal-to-noise ratio of our setup.

In order to do a quantitative analysis of the data, we fit the time evolution of the asymmetry for both energy intervals using a double-exponential function convolved with a Gaussian function G(t) with a full width at half maximum of 58~fs:
\begin{eqnarray*}
    \frac{A(t)}{A(t<t_0)} &=& G(t) \circledast  \Bigg[ 1-\Theta(t-t_0)\cdot \bigg[ \bigg. \Delta A_m \cdot \left( 1- e^{- \frac{t-t_0}{\tau_m}}\right) \\ &\,&- \Delta A_r \cdot \left(1-e^{- \frac{t-t_0}{\tau_r}} \right) \bigg] \Bigg. \Bigg].
    \label{time_dependent}
\end{eqnarray*}
Here, $t_0$ defines the onset of the demagnetization process, $\tau_m$ and $\tau_r$ are the demagnetization and remagnetization constants, respectively, $\Delta A_m$ is proportional to the maximal fractional demagnetization, and $\Delta A_r$ is proportional to the fast remagnetization. For our data, $\Delta A_r$ is smaller than $\Delta A_m$, meaning that the sample does not return to its ground state in the measured 10-ps time range. Full remagnetization is only achieved on a longer timescale. The least-square fit results are listed in table \ref{tab:fit_results_double}. In addition to the delayed dynamics in nickel, we observe a faster demagnetization time at the nickel M-edge as well as a slightly stronger demagnetization for the selected spectral ranges.

\begin{table}
\begin{ruledtabular}
\begin{tabular}{ccc}
Fe, 50.7 to 54.5 eV  & 1 hour & 12 hours\\\hline
$t_ 0$~(fs)  & $ 7 \pm 10 $ & $ 0 \pm 3$ \\ 
$\Delta A_m$ & $ 0.253 \pm 0.012 $ & $ 0.276 \pm 0.005 $ \\
$\Delta A_r$ & $ 0.182 \pm 0.010 $ & $ 0.199 \pm 0.004 $ \\
$\tau_m$~(fs)& $ 185\pm 21 $ & $ 214 \pm 8 $\\
$\tau_r$~(ps)& $ 1.96\pm 0.24 $ & $ 1.85 \pm 0.08$\\\hline\hline

 Ni, 63.4 to 67.5 eV & 1 hour & 12 hours\\\hline
$t_ 0$~(fs) & $ 67 \pm 5 $ & $ 55.2 \pm 1.9$ \\ 
$\Delta A_m$ & $ 0.264 \pm 0.008 $ & $ 0.279 \pm 0.003 $ \\
$\Delta A_r$ & $ 0.197 \pm 0.007 $ & $ 0.206 \pm 0.003 $ \\
$\tau_m$~(fs)& $ 137 \pm 10 $ & $ 147 \pm 4 $\\
$\tau_r$~(ps)& $ 1.5 \pm 0.1 $ & $ 1.67 \pm 0.04 $\\
\end{tabular}
\end{ruledtabular}
\caption{\label{tab:fit_results_double} Fit results using Eq.~\ref{time_dependent} of the demagnetization traces in Fig.~\ref{fig:FeNi_tr}.  For the least-squares fit, measurement data were weighted according to the data spread at each delay individually. The given error margins correspond to the 1-sigma standard deviations of the fit.}
\end{table}

Crucially, the fit results for the 1-hour data set differ only slightly from the results for the complete 12-hour data. The differences in demagnetization dynamics between iron and nickel are well-reproduced by this much shorter measurement. To verify that these results are not simply by choice of our 1 hour data set, we split the 12-hour data set in one-hour chunks and repeated the least-square fit for each of these. From this analysis, we find that generally, the 1-hour fit results closely match with the 12-hour result. However, the predicted error margins from a single least-square fit underestimate the true variance of the analysis results by roughly a factor two. For example for the onset of demagnetization in nickel, we find values centered on 53~fs with a 1-sigma standard deviation of 9~fs. Such differences are acceptably small however, and can easily be explained by small drifts in the experimental setup over the course of the long measurement. 

\section{\label{sec:conclusions}Conclusion}
In summary, we have presented a novel setup combining a high-repetition-rate bright high harmonic light source with a T-MOKE-setup enabling high magnetic fields up to 0.86~T and sample temperatures down to 10~K. We have shown that this setup allows for the detailed measurement of element-specific magnetization behavior in complex samples such as the double perovskite LNMO. Furthermore, using a Fe$_{19}$Ni$_{81}$ permalloy thin film as a well-known reference sample, we have demonstrated the capabilities of our setup to measure high-quality ultrafast demagnetization dynamics. We find that already a single hour of measurement time allows for a detailed view into the ultrafast dynamics in a 10-picosecond time window. For example, this measurement already allows us to determine the relative delay of the demagnetization in nickel compared to iron with 10~fs accuracy. Extending our measurement to 12 hours, we acquire excellent data with very low noise, which can be directly attributed to the use of the high-power Yb-fiber laser, operating at 100 to 300~kHz repetition rate. 

In the future, we will apply our setup to the measurement of element-resolved magnetic responses as a function of temperature, time and magnetic field. Amongst others, this work will contribute to a more profound microscopic understanding of the dynamics in complex magnetic alloys and strongly correlated electron systems.

\section{ACKNOWLEDGEMENTS}
H.P, K.S, S.M., V.M. and D.S. acknowledge support from the German Science Foundation through project number 399572199. G.S.M.J. acknowledges funding by the Alexander von Humboldt Foundation. We thank Manfred Albrecht from the University of Augsburg for the fabrication of the Permalloy sample. Also, we wish to thank Christof Schmidt and the team of the central workshop of physics for their excellent support in constructing the HHG EUV-T-MOKE setup. 

\section{Data availability}
The data that support the findings of this study are available from the corresponding author upon reasonable request.

\section{Sample Preparation}
The double perovskite LNMO film with a thickness of $d\approx100$~nm was epitaxially grown on a MgO(100) substrate by means of a solution-based and vacuum-free metalorganic aerosol deposition (MAD) technique~\cite{jungbauer2014atomic}.

The investigated Fe$_{19}$Ni$_{81}$ sample is a 15-nm thin film deposited with magnetron sputtering at room temperature on a silicon substrate (Si/SiO$_{2}$(100~nm)) and capped with 5~nm Si$_{3}$N$_{4}$. The sample is poly-crystalline and exhibits an in-plane magnetization orientation. 

\section{REFERENCES}
\bibliography{bib_HHG_TMOKE}% Produces the bibliography via BibTeX.
\end{document}